# Towards an Ontology-Driven Blockchain Design for Supply Chain Provenance

Henry M. Kim
Schulich School of Business, York University, 4700 Keele St.., Toronto, Ontario Canada
hmkim@yorku.ca | 416-736-2100

Marek Laskowski
York University, 4700 Keele St.., Toronto, Ontario Canada
marlas@yorku.ca | 416-736-2100

## Abstract

An interesting research problem in our age of Big Data is that of determining provenance. Granular evaluation of provenance of physical goods--e.g. tracking ingredients of a pharmaceutical or demonstrating authenticity of luxury goods--has often not been possible with today's items that are produced and transported in complex, inter-organizational, often internationally-spanning supply chains. Recent adoption of Internet of Things and Blockchain technologies give promise at better supply chain provenance. We are particularly interested in the blockchain as many favoured use cases of blockchain are for provenance tracking. We are also interested in applying ontologies as there has been some work done on knowledge provenance, traceability, and food provenance using ontologies. In this paper, we make a case for why ontologies can contribute to blockchain design. To support this case, we analyze a traceability ontology and translate some of its representations to smart contracts that execute a provenance trace and enforce traceability constraints on the Ethereum blockchain platform.

# Towards an Ontology-Driven Blockchain Design for Supply Chain Provenance

## Introduction

An interesting practical and theoretical problem in our age of Big Data is that of determining source of information. One community of researchers interested in addressing this problem is the ontological engineering community, who are actively researching the development of ontologies for knowledge provenance (Fox and Huang 2005)(Erickson et al. 2016).

According to Merriam-Webster, provenance is "source or origin; or, the history of ownership of a valued object or work of art or literature" (Merriam-Webster 2016). The ontological engineering community's efforts at formally representing and reasoning about the provenance of knowledge on the Web can be considered tractable because data required to ascertain provenance is in digital form—as data, meta-data, and timestamps, for example. Moreover, semantic Web technologies facilitate the semantic and workflow modelling and inference required for Web knowledge provenance. Arguably, provenance evaluation of artifacts that do not have such a ready and openly accessible digital footprint or facilitating technologies has not been as tractable a problem to address. Tracking the ingredients of a pharmaceutical or demonstrating authenticity of a luxury handbag are some examples. Whereas it is true that UPS can accurately track its packages, such granular provenance evaluation has often not been possible with today's items that are produced and transported in complex, inter-organizational, often internationally-spanning supply chains.

As of late, however, new technologies, namely Internet of Things (IoT) and Blockchain technologies, promise to offer provenance even in complex supply chains (Armstrong 2016). Internet-aware sensors capture finely granular real-time data about product and environment characteristics as well as location and timestamps throughout the supply chain. So lack of a digital footprint may no longer be an issue. Furthermore, distributed, shared databases using Blockchain technologies promise to offer highly secure and immutable access to supply chain data. Blockchain databases are decentralized so that provenance can be evaluated even when no one party can claim ownership over all supply chain data. Inasmuch as metadata and semantic Web technologies enabled ontologies to be applied for knowledge provenance, it is a key premise of our research that IoT and the Blockchain, in particular, now can enable ontologies to be used for much improved supply chain provenance. Armed with this premise, this paper details our efforts towards developing an ontology-based blockchain for supply chain provenance.

The paper then is organized as follows. In next section, we expound the Blockchain, which constitutes the enabling technology for our work. Excerpts of the TOVE Traceability Ontology which serves as the ontology source for our blockchain are presented next. Following this, a proof-of-concept implementation of a provenance evaluating blockchain executed on the *Ethereum* application development platform and encoded in the *Solidity* language is presented. Finally, we present concluding remarks and commentary for future work.

## The Blockchain

A blockchain is "a distributed database that maintains a continuously-growing list of data records secured from tampering and revision. It consists of blocks, holding batches of individual transactions. Each block contains a timestamp and a link to a previous block" (Morris 2016; Nakamoto 2008; Popper 2016). This cryptographic technology "offers a way for people who do not know or trust each other to create a record of who owns what that will compel the assent of everyone concerned. It is a way of making and preserving truths" (The Economist Staff 2015).

Originally developed to underpin the bitcoin cryptocurrency network, the blockchain has many enthusiastic supporters who see its potential beyond cash and currency (Boroujerdi and Wolf 2015). The potential for blockchain to enable a distributed ledger of digital assets is the source of their enthusiasm (Tapscott and Tapscott 2016, p. 7):

*Some scholars have argued that the invention of double-entry bookkeeping enabled the rise of capitalism and the nation-state. This new digital ledger of economic transactions can be programmed to record virtually everything of value and importance to humankind: birth and death certificates, marriage licenses, deeds and titles of ownership, educational degrees, financial accounts, medical procedures, insurance claims, votes, provenance of food, and anything else that can be expressed in code.*

A more circumspect perspective on the potential for blockchain views the following as "genuine" blockchain use cases: 1) inter-organizational recordkeeping, 2) lightweight financial systems such as crowdfunding, gift cards, and loyalty points, 3) multiparty aggregation to address the infrastructure difficulty of combining information from large number of sources, and 4) provenance tracking (Greenspan 2016). As it is explicated by both this and Tapscotts' perspectives, it seems that provenance tracking along a supply chain could be one of the killer apps of blockchain. Already there are startups like provenance.org and skuchain that are exploring this possibility. We believe that works from the computational ontology research community can be useful for these startups and other researchers interested in this topic. That is, *specifically*, we believe ontologies can contribute to develop blockchain applications for supply chain provenance. In fact, *in general*, we believe that ontologies can contribute to developing blockchain applications.

*Why Use Ontologies for Blockchain Development?*

For the general case, recall this: at its heart, a blockchain is a distributed database. In order to understand data in a database distributed across numerous organizations, there must be common interpretation of data across these organizations. This interpretation can be informally enforced via use of common data standards—i.e. models, dictionaries, and conventions—and via business practices and processes that support adoption of data standards by human developers working at these organizations. Interpretation can also be formally enforced via formal specifications that enable automated inference and verification within software applications that execute on a network that spans these organizations.

Concomitantly, the classic definition of a computational ontology (Gruber 1993) is that it is "an explicit specification of a conceptualization." In ontology-based enterprise modeling, the conceptualization is the set of ontologies required to ensure common interpretation of data from one or more enterprises' shared databases. Such ontologies can be informal or light-weight (e.g. North American Industry Classification System [NAICS]); formal, like the TOVE Ontologies (Fox and Gruninger 1998); or be somewhere in between (i.e. semi-formal). Making the reasonable assumption that blockchain modeling is a specialized form of inter-enterprise modeling, we make the case that ontology-based blockchain modeling will result in a blockchain with enhanced interpretability. That is:

- A modeling approach based on informal or semi-formal ontologies can lead to better data standards, and business practices and processes for developing and operating a blockchain.
- A modeling approach based on formal ontologies can aid in the formal specifications for automated inference and verification in the operation of a blockchain.

It is this latter point that is particularly interesting because that description is very similar to the definition of *smart contracts* as "pieces of software that represent a business arrangement and execute themselves automatically under pre-determined circumstances" (The Economist Staff 2016). Smart Contracts are critical for widespread blockchain adoption. Arguably, the second and third largest blockchain endeavours are Ethereum and R3CEV, with the first clearly being bitcoin. Ethereum is a worldwide platform for implementing distributed applications. It is run on a public, permission-less blockchain upon which smart contracts are executed. Ethers represent Ethereum's crypto-currency paid to participants who maintain the blockchain; there are $1.1B USD equivalent of ethers in circulation[1]. R3CEV is a startup funded for $200M USD[2] by a consortium of over 40 financial institutions worldwide. Its main focus has been to develop smart contracts that access block-chained distributed ledgers of consortium institutions to, for example, automatically execute terms of interest rate swaps between two banks (Rizzo 2016).

Given the importance placed on smart contracts, the following modified statement outlines a very compelling rationale for ontology-based blockchains:

- A modeling approach based on formal ontologies can aid in the formal specifications for automated inference and verification in the operation of a blockchain. *That is, a modeling approach based on formal ontologies can aid in the development of smart contracts that execute on the blockchain.*

Now that we have a made a general case for developing ontology-based blockchains, we make the specific case: In the next section, we outline the TOVE Traceability Ontology as an apt source for our ontology-based blockchain for supply chain provenance, and present excerpts relevant to develop a proof-of-concept.

---

[1] According to coinmarketcap.com, as of August 22, 2016. In contrast, there are $9.2B USD equivalent bitcoins in circulation
[2] https://www.cryptocoinsnews.com/report-blockchain-r3-seeks-200-million-backers/

**Traceability Ontology Based Blockchain for Provenance**

As mentioned, blockchain startups like provenance.org and skuchain are working on supply chain provenance. However, there does not appear to be any works other than our own effort at taking an ontology-based approach. There have been some related works using ontologies, though not blockchain related. Arguably, the most expansive is the traceability ontology (Kim et al. 1995) that served as a key part of TOVE Ontologies for enterprise modelling (Fox and Gruninger 1998). This traceability work has garnered interest, interestingly, from food sciences researchers (Dabbene et al. 2014)(Regattieri et al. 2007). Food science has co-opted what was an ontology biased towards manufacturing industry enterprise modeling to ensure food safety along the food supply chain. This bodes well for using the TOVE Traceability Ontology as the primary source to design our blockchain.

*TOVE Traceability Ontology Excerpt*

Here are the key informal assumptions used in developing this ontology

- It must be possible to trace from one entity to another, where neither the entities are abstracted entities.
- *Traceable Resource Unit* (aka a *TRU*—a representation for a batch of a something, e.g. a tru of 100 widgets) is the resource representation that must be traceable, since a tru is neither an abstracted nor aggregated entity.
- Primitive activity is the activity representation that must be traceable, since a primitive activity is neither an abstracted nor aggregated entity.

The following is a simplified version of the data model for the ontology (Kim et al. 1995).

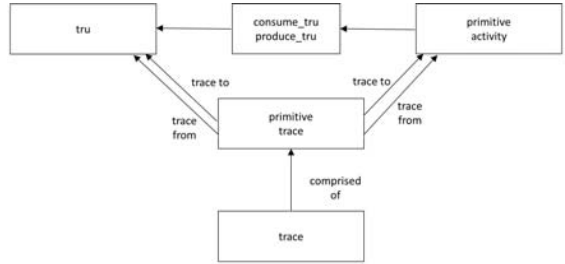

Figure 1: Simplified TOVE Traceability Ontology Data Model

The following are some key axioms of the ontology expressed formally in first-order logic (Kim 1999):

Trace Axiom: Cons-1. **A tru is produced only once.**

$$\forall A \forall St_1 \forall Rt \forall s \, [ \, holds(produce(St_1,A),s) \land holds(produces(St_1,Rt),s) \supset \\ \neg \exists St_2 \, \{ \, holds(produce(St_2,A),s) \land holds(produces(St_2,Rt),s) \land St_1 \neq St_2 \, \} \, ].$$

Rt: a tru
$St_1, St_2$: the same state describing the production of Rt
A: an activity which produces Rt
s: an extant situation

T4. If a tru is used or consumed by a primitive activity, then the resource point of the tru at time $T_p$ (the time at which the use or consume state is enabled) is the amount of the tru that has not been committed, and hence is available for other states.

Figure 2: Key Axioms of the TOVE Traceability Ontology

## Proof-of-Concept Implementation of Ontology Based Blockchain for Supply Chain Provenance

In this section we describe how we interpret the traceability ontology as a real-time tracking system, capable of tracing the provenance of TRU's back to any other TRU's in their provenance history or chain.

As suggested by Figure 3, Blockchain technologies are built upon Internet technologies, using a Web browser as a natural interface. We used The Truffle framework by ConsenSys[3] to generate a JavaScript-based (Web3 ABI) interface to interact with the deployed smart contract, forming inputs, or predicates, into the system to define the state of objects, as well as performing traces.

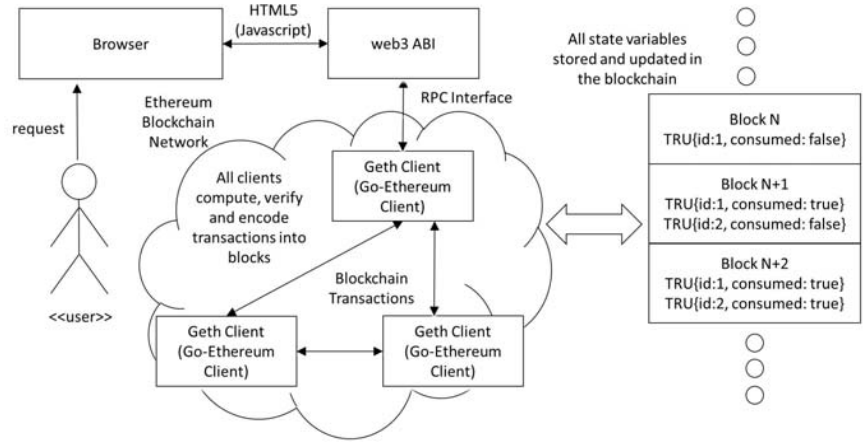

**Figure 3: A system diagram depicting mediated user interaction with the Blockchain application**

The state of the Blockchain or distributed ledger in Ethereum represents the state of all deployed programs, or contracts, in terms of inputs, internal variables, and outputs (e.g. logs). All Ethereum clients on the network can participate in maintaining the ledger by listening for, computing, verifying and encoding transactions into blocks (i.e Mining). Solidity is currently the main programming language on the Ethereum platform, and it is purpose-built for writing smart contract style programs. Solidity is an Object Oriented (OO) language, in which the Contract is the fundamental class for encapsulating programs or smart contracts in Solidity. While the language and platform are representationally Turing Complete—they can be used to represent any possible computation—in practice, computations within Contracts are subject to constraints. These are in turn due to the economic incentives used to reward the decentralized network of individuals who carry out computations on the blockchain in order to determine its next state, or block. That is, that all transactions have a cost that has to be paid in Ether, Ethereum's native token-based currency.

---

[3] https://consensys.net/

Data Models such as Figure I used in ontology-based enterprise modelling and their subsequent implementation in Object Oriented Programming environments have been extensively explored in the literature (Evermann and Wand 2005)(Siricharoen 2007). One such methodology commonly used as an intermediary representation for translating business processes into the language of Object Oriented Software Engineering is UML (Eriksson and Penker 2000).

Here then is a pertinent diagram.

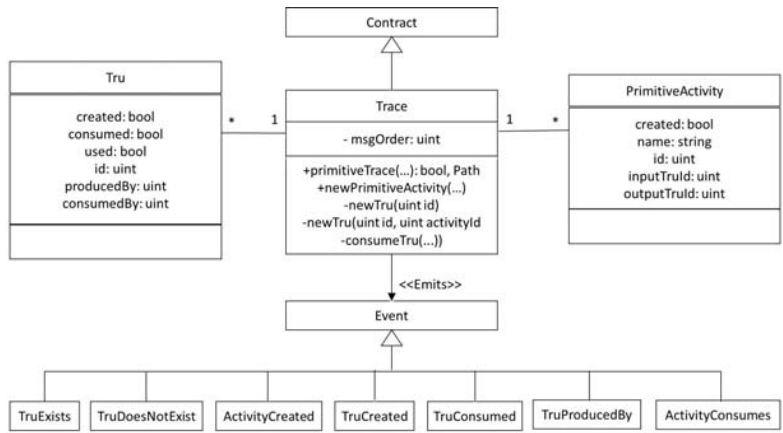

**Figure 4: A Hybrid UML Diagram Depicting the Object Oriented Design of the Traceability Data Model, as Implemented in Ethereum-Solidity**

As implied by Figure 4, the only public interfaces are provided by the `Trace` class, therefore all user input affecting the contract state on the blockchain is through the `Trace` class. All output communication from the Contract is accomplished through the use of Events, which are log data variables to the blockchain and in turn read by the Ethereum client. Ensuring that constraints and relationships implied by the axioms are applied to the system is analogous to maintaining the so-called class invariants within the system of objects (Meyer 1988).

Translating from formal ontology representations to Solidity can be problematic due to Solidity's novelty and correspondingly low maturity. Therefore, as shown in Figure 4, the `Trace` object or class is represented as a `Contract`, and the `PrimitiveActivity` and `Tru` are represented as struct types which are essentially classes without associated methods, or functions bound to each object instance. `PrimitiveActivity` and `Tru` could also be implemented as contracts. However, the semantics and implications of doing so would considerably complicate the discussion in this paper, without any apparent benefit. As noted in the listings and UML diagram, `Trace` has several functions defined, which encompass the behavior and constraints upon the encapsulated types, `PrimitiveActivity` and `Tru`. The public functions comprise the public interface for the contract. The notion of a `Primitive Trace` is thereby implemented as a public function or method rather than an object or variable, which is a common approach to implementing computed fields or variables bound to objects (Meyer 1988).

The Appendix shows partial listing of the source code for the smart contracts we implemented. The full version is available for download here: https://github.com/professormarek/traceability. The gist of the code is that we are able to record the scenario pictured in Figure 5 to the Ethereum blockchain, and the smart contract implementation of the ontology axioms are used to generate the trace shown in Figure 6.

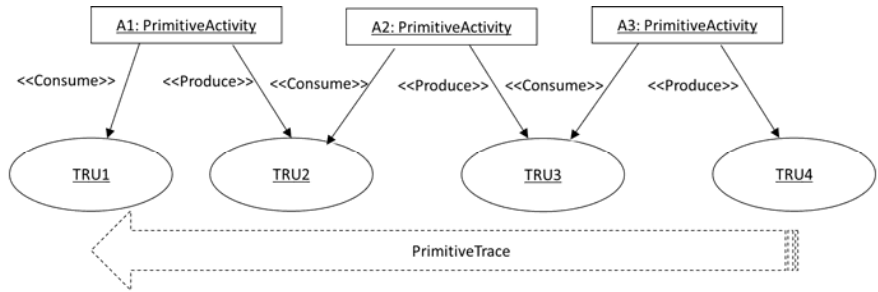

**Figure 5: Trace Scenario considered for Proof of Concept Demonstration**

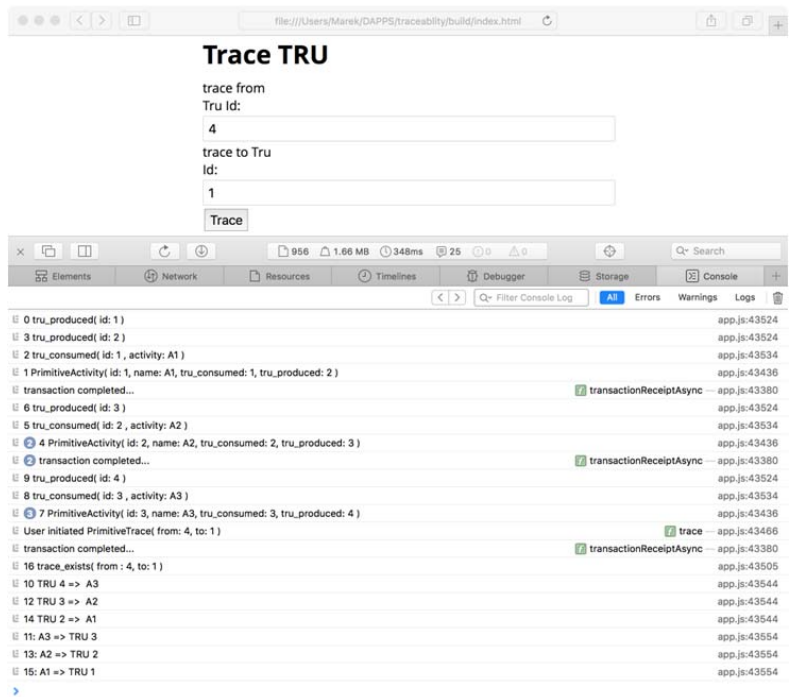

**Figure 6: Screen Output of a Trace Executed on the Ethereum Blockchain**

## Concluding Remarks and Future Work

We identified evaluating provenance as an important and ongoing business issue. Evaluating knowledge provenance has become more possible as more and more of the data required to discover the source of knowledge is recorded on the Web. Evaluating provenance of physical goods—or what we call supply chain provenance—has generally been more difficult because so many goods are handled in complex, international supply chains where granular tracking of physical characteristics and product whereabouts has not been possible. That is, until recently, when provenance evaluation has become more possible with the advent of IoT and Blockchain.

In particular, as blockchain technology evolves, as more business models that leverage it are conceived, and as more researchers explore still-nascent research opportunities with its use, we believe that the ontological engineering community can make a contribution to the growth of blockchain. We posit one specific and two general potential contributions and present preliminary results in this paper as a proof-of-concept of these contributions.

- Specifically, ontologies that represent fundamental concepts in traceability can contribute domain knowledge to develop blockchain applications for supply chain provenance. As a proof-of-concept, we wrote source code on the Ethereum blockchain and assessed that we could in fact program concepts from the TOVE Traceability Ontology in a blockchain platform.
- Generally, a modeling approach based on informal or semi-formal ontologies can lead to better data standards, and business practices and processes for developing and operating a blockchain. As a proof-of-concept, we analyzed excerpted assumptions and data models of the TOVE Traceability Ontology and used them to develop the appropriate distributed ledger on the blockchain.
- Generally, a modeling approach based on formal ontologies can aid in the development of smart contracts that execute on the blockchain. As a proof-of-concept, we translated TOVE Traceabiliy Ontology axioms that were expressed in first-order logic into smart contracts that could execute a provenance trace and enforce traceability constraints on the blockchain.

There is much more work to be done. Specifically, there are many more traceability constructs—both informal data models and formal axioms—that ought to be encoded to enhance blockchain provenance capabilities. Generally, more research is needed to make the conversion from ontology representations to blockchain code more systematic. That may entail more granularly outlining conversion steps, developing custom API's, or contributing to efforts to convert semantic Web representations like OWL and RDF into blockchain-compliant representations.

Future work notwithstanding, we believe that we have already made some contribution towards providing guidance for those wishing to use ontologies to develop blockchain applications, and more specifically, for evaluating supply chain provenance.

# Appendix: Partial Listing of the Trace Construct[4]

```
1  contract Trace{
2      struct Tru{
3          bool consumed;
4          bool used;
5          bool created;
6          uint id;
7          uint producedBy;
8          uint consumedBy;
9      }
10     struct PrimitiveActivity{
11         bool created;
12         string name;
13         uint id;
14         uint inputTruId;
15         uint outputTruId;
16     }
17     mapping(uint => Tru) truLookup;
18     mapping(uint => PrimitiveActivity) activityLookup;
19     uint msgOrder;
20     function Trace(){
21         msgOrder = 0;
22     }
23     modifier nonZero(uint num){
24         if(num == 0){
25             throw;
26         }
27         _
28     }
29     modifier truDoesNotExist(uint id){
30         if(truLookup[id].created){
31             throw;
32         }
33         _
34     }
35     modifier truAvailable(uint id){
36         if(truLookup[id].consumed || truLookup[id].used){
37             throw;
38         }
39         _
40     }
41     modifier truExists(uint id){
42         if(truLookup[id].created != true){
43             throw;
44         }
45         _
46     }
47     modifier primitiveActivityExists(uint id){
48         if(activityLookup[id].created != true){
49             throw;
50         }
51         _
52     }
53     function newTru(uint id) private
54     truDoesNotExist(id)
55     nonZero(id)
56     {
57         truLookup[id].created = true;
58         truLookup[id].id = id;
59         truLookup[id].consumed = false;
60         truLookup[id].used = false;
61         truLookup[id].producedBy = 0;
62         truLookup[id].consumedBy = 0;
63         TruCreated(msgOrder++, id);
64     }
65     function newTru(uint id, uint activityId) private
66     truDoesNotExist(id)
67     nonZero(id)
68     primitiveActivityExists(activityId)
69     {
70         newTru(id);
71         truLookup[id].producedBy = activityId;
72     }
73     function consumeTru(uint truId, uint activityId) private
74     truExists(truId)
75     truAvailable(truId)
76     primitiveActivityExists(activityId)
77     {
78         truLookup[truId].consumed = true;
79         truLookup[truId].consumedBy = activityId;
80         TruConsumed(msgOrder++, truId,
               activityId,
               activityLookup[activityId].
               name);
81     }
```

---